\documentstyle[preprint,prd,aps]{revtex}   
\begin{document} 
\draft
\title{Tracker fields from nonminimally coupled theory}
\author{Ruggiero de Ritis}
\address{Dipartimento di Scienze
Fisiche, Universit\`a di Napoli, Mostra d'Oltremare Pad. 20, I-80125
Napoli, Italy \\   and Istituto Nazionale di Fisica Nucleare, Sez. 
Napoli, Monte Sant'Angelo, Via Cinzia, Ed.  G, I-80126 Napoli, Italy}
\author{Alma A. Marino} 
\address{Osservatorio Astronomico di
Capodimonte, Via Moiariello 16, I-80131 Napoli, Italy \\ and Istituto
Nazionale di Fisica Nucleare, Sez. Napoli, Monte Sant'Angelo, Via
Cinzia, Ed. G, I-80126 Napoli, Italy}  
\author{Claudio Rubano}
\address{Dipartimento di Scienze Fisiche,
Universit\`a di Napoli, Mostra d'Oltremare Pad. 20, I-80125 Napoli,
Italy \\  and Istituto Nazionale di Fisica Nucleare, Sez. Napoli,
Monte Sant'Angelo, Via Cinzia, Ed. G, I-80126 Napoli, Italy}
\author{Paolo Scudellaro}
\address{Dipartimento di Scienze Fisiche,
Universit\`a di Napoli, Mostra d'Oltremare Pad. 20, I-80125 Napoli,
Italy \\ and Istituto Nazionale di Fisica Nucleare, Sez. Napoli,
Monte Sant'Angelo, Via Cinzia, Ed. G, I-80126 Napoli, Italy}
\date{\today}
\maketitle
\begin{abstract}
We extend the concept of quintessence to flat nonminimally coupled
scalar--tensor theories of gravity. By means of Noether's symmetries
for the cosmological pointlike Lagrangian $\cal L$, it is possible
to exhibit exact solutions for a class of models depending on a free
parameter $s$. This parameter comes out in the relationship existing
between the coupling $F(\varphi)$ and the potential $V(\varphi)$
because of such a symmetry for $\cal L$. When inverse power--law
potentials are taken into account, a whole family of exact solutions
parametrized by such an $s$ is proposed as a class of tracker fields,
and some considerations are made about them.  
\end{abstract}

\narrowtext
\section{Introduction}

Recently, astronomical observations have indicated a strong evidence
of an accelerated universe \cite{Perl1,Perl2,Leib}. Together with
measurements of the cosmic microwave background and the mass power
spectrum (see \cite{Wang}, for example), they suggest that a large
amount of the energy density of the universe should have a negative
pressure. A way of describing the {\it missing} component of the
energy needed to reach, for instance, the critical energy density is
{\it quintessence} \cite{Ostrik,CDS,ZWS,Wang}. Essentially, this is
a spatially inhomogeneous and slowly evolving fraction of the total
energy. We may consider it as given by a scalar field $\varphi$
slowly rolling down its potential $V(\varphi)$ and such that $-1 <
w_{\varphi} < 0$, being $w_{\varphi} \equiv
p_{\varphi}/{\rho_{\varphi}}$, where $p_{\varphi}$ and
$\rho_{\varphi}$ are, respectively, the pressure and the energy
density of the scalar field. Actually, recent considerations
\cite{Perl3} fix the interval $-1 < w_{\varphi} \lesssim -0.6$ as the
more suitable one for such a scalar field to effectively represent
quintessence. (A cosmological constant $\Lambda$, which mimics
vacuum energy density, also produces a negative pressure, but this
is such that $p_{\Lambda}/{\rho_{\Lambda}} = -1$.)

Within the scenario created by quintessence, there is a twofold
problem. One of its two aspects is the so--called {\it fine--tuning
problem}, based on the question why $\rho_{\varphi}$ appears to be
so small with respect to typical particle physics scales. The other
aspect, called {\it cosmic coincidence} \cite{Stein}, requires that
the initial conditions have to be set precisely in order to explain
why $\rho_{\varphi}$ and the matter energy density $\rho_m$ should
appear of the same order today. This poses problems on the
theoretical choice for the energy fraction which seems to be missing.

More recently, a special form of quintessence has been introduced to
avoid the coincidence problem. It is called {\it tracker field}
\cite{ZWS,SWZ} and works like an attractor solution to the equations
of motion, even if it is not really a fixed point, since it is time
dependent and $\rho_{\varphi}/{\rho_m}$ changes as $\varphi$
evolves, leaving later cosmology independent of the early
conditions. Of the kinds of potentials proposed for quintessence, two
have been more studied for tracker solutions, namely, $V(\varphi) =
M^{4 + \alpha} {\varphi}^{-\alpha}$ and $V(\varphi) = M^4
[exp({M_P}/{\varphi}) - 1]$, where $M$ and $\alpha > 0$ are free
parameters, and $M_P$ is the Planck mass. The family of tracker
solutions is then parametrized by $M$, whose value can be fixed by the
measured value of $\Omega_m$ today. Such specific forms of potentials
have been chosen because of their importance in particle physics models
\cite{Aff,Barrow,Barreiro,Bin}. Anyway, to our knowledge, Ratra and
Peebles \cite{Peebles,Ratra} were the first ones to study the
influence of an inverse power--law potential in cosmology with a
scalar field. 

A general study of specific features of tracker solutions has been
made \cite{ZWS,SWZ}, also leading to the introduction of the
important function $\Gamma$, in order to fulfill the so--called {\it
tracker equation}. Essentially, it is shown that ``tracking behavior
with $w_{Q} < w_{B}$ occurs for any potential in which $\Gamma
\equiv {V''}V/(V')^2 > 1$ and is nearly constant'' for any possible
initial value of the scalar field $Q$. (Here, prime denotes
derivative with respect to $Q$, and $B$ indicates {\it background}.)
As a consequence, once a certain potential has been assigned, the
existence of tracking solutions can be tested without solving the
equations of motion.

Usually, the scalar field representing quintessence has been
considered as minimally coupled to gravity and only more recently
nonminimal coupling has been introduced in such a context
\cite{Uzan,Amendola,PBM}. In this connection, we consider interesting
to refer again to scalar--tensor theories of gravity, in which the
scalar field $\varphi$ is nonminimally coupled to gravity and also
inverse power--law potentials for $\varphi$ have been studied,
leading to exact solutions for $\varphi (t)$ and for the scale factor
$a(t)$ of the universe (see \cite{RivNC} and references therein for
a review on this topic). Our main purpose is to outline how a family
of exact tracker solutions can be derived from a flat nonminimally
coupled theory with inverse power--law potential, expliciting also
how the tracker equation is fulfilled in such a way.

In what follows, section $2$ is devoted to a short review of some
basic notions of flat nonminimally coupled theories, and section $3$
selects a special class of solutions. In section $4$ we identify such
a class as a family of tracker solutions, and in section $5$ we draw
conclusions. 

\section{Nonminimally coupled theories}

As it is well known, one of the main reasons why nonminimally coupled
scalar--tensor theories of gravity have received a special degree of
attention is that they seem to play an important role in inflationary
cosmology (see \cite{RivNC}, for instance). In the following, however,
we will limit ourselves to concentrate on what can be useful here,
trying to draw a sort of narrow and straightforward path to our
specific goal, namely deriving tracker solutions. First of all, we
deal with flat (i.e., with the curvature scalar $k = 0$) models
described by the action 

\begin{equation}
{\cal A} = \int {d^{4}x \sqrt{-g} \left[ F(\varphi)R + \case1/2
g^{\mu\nu} {\varphi}_{;\mu} {\varphi}_{;\nu} - V(\varphi) + {\cal L}_m
\right] } \,, 
\label{eq1}
\end{equation}
where g is the determinant of the metric $g_{\mu \nu}$, $R$ the
curvature scalar, semicolon indicates ccovariant derivative, and the
functions $F(\varphi)$ and $V(\varphi)$ are not specified; ${\cal
L}_m$ is the Lagrangian of an ordinary perfect fluid noncoupled to the
scalar field $\varphi$. $F(\varphi)$ expresses the nonminimal coupling
of $\varphi$ with gravity and is such that, when $F \equiv F_{0} \equiv
- 1/2$ (using units such that $8{\pi}G = c = \hbar = 1$), action in
Eq.\ (\ref{eq1}) reduces to the usual one in the flat minimal coupling
case.

The field equations can be derived by varying Eq.\ (\ref{eq1}) with
respect to $g_{\mu\nu}$ and they can be written as 

\begin{equation}
G_{\mu \nu} = \tilde{T}_{\mu \nu} \equiv - \frac{1}{2F(\varphi)}
T^{(tot)}_{\mu \nu} \,, 
\label{eq2} 
\end{equation}
where $G_{\mu \nu} \equiv R_{\mu \nu} - {\case1/2}g_{\mu \nu}R$ is the
Einstein tensor, and $\tilde{T}_{\mu \nu}$ is a quantity related to the
total stress--energy tensor 

\begin{equation}
T^{(tot)}_{\mu \nu} \equiv T^{(\varphi)}_{\mu \nu} + T^{(m)}_{\mu \nu}
\,.  
\label{eq3} 
\end{equation}
Here, the tensor

\begin{eqnarray} 
T^{(\varphi)}_{\mu \nu} \equiv {\varphi}_{;\mu}{\varphi}_{;\nu} -
\frac{1}{2} g_{\mu \nu}{\varphi}_{;\alpha}{\varphi}^{;\alpha}
+ g_{\mu \nu}V(\varphi) \nonumber \\
+ 2g_{\mu \nu}\Box F(\varphi) - 2F(\varphi)_{;\mu \nu}  
\label{eq4}
\end{eqnarray} 
represents the scalar field source, while $T^{(m)}_{\mu \nu}$ is the
standard perfect fluid matter source (and $\Box$ is the usual
d'Alembert operator). Varying with respect to $\varphi$, we get the
Klein--Gordon equation ruling the dynamics of the scalar field
$\varphi$ 

\begin{equation}
\Box \varphi - RF'(\varphi) + V'(\varphi) = 0 \,,
\label{eq5}
\end{equation}
denoting the prime the derivative with respect to $\varphi$. It is
possible to show that Eq.\ (\ref{eq5}) is nothing else but the
contracted Bianchi identity \cite{CR1,RivNC}, which means that the
effective stress--energy tensor $\tilde{T}_{\mu \nu}$ introduced in
Eq.\ (\ref{eq2}) is a zero--divergence tensor, coherently with
Einstein's theory of gravity \cite{Madsen}.

Fixing a homogeneous and isotropic (FRW) metric reduces the relevant
variables to $a$ and $\varphi$, i.e. the scale factor and the scalar
field, each one a function of $t$ only. As a matter of fact, field
equations \ (\ref{eq2}) can be reduced to two ordinary differential
equations ($k = 0$)
 
\begin{equation}
H^{2} + \frac{\dot{F}}{F} H + \frac{{\rho}_{\varphi}}{6F} +
\frac{{\rho}_{m}}{6F} = 0 \,,
\label{eq6}
\end{equation}

\begin{equation}
\dot{H} = \frac{{\dot{\varphi}}^2}{4F} - \frac{1}{2} \left(H^2 +
\frac{{\rho}_{\varphi}}{6F} \right) - \frac{\ddot{F}}{2F} +
\frac{p_m}{4F} + \frac{\rho_m}{6F} \,,
\label{eq7}
\end{equation}
where dot indicates the time derivative, $H \equiv \dot{a}/a$, $p_m =
w_{m}{\rho}_{m}$ is the equation of state for ordinary fluid matter,
and $p_{\varphi} = w_{\varphi}{\rho}_{\varphi}$ is the equation of
state for the scalar field $\varphi$, having defined its pressure
and energy density, respectively, as

\begin{equation}
p_{\varphi} = {\case1/2}{\dot{\varphi}}^2 - V(\varphi) \,,
\label{eq8a}
\end{equation}

\begin{equation}
{\rho}_{\varphi} = {\case1/2}{\dot{\varphi}}^2 + V(\varphi) \,.
\label{eq8b} 
\end{equation}
This implies, thus, that

\begin{equation}
w_{\varphi} = \frac{p_{\varphi}}{{\rho}_{\varphi}} =
\frac{{\dot{\varphi}}^2 - 2V(\varphi)}{{\dot{\varphi}}^2 +
2V(\varphi)} \,.
\label{eq9}
\end{equation}

Now, it is very interesting, and for us very important, to notice
that Eq.\ (\ref{eq5}) (rewritten in the FRW flat case) and Eq.\
(\ref{eq7}) can be seen as the Euler--Lagrange equations of the point
Lagrangian 

\begin{equation} 
{\cal L} = 6a{\dot{a}}^{2} F(\varphi) + 6{\dot{a}}{\dot{\varphi}}a^{2}
F'(\varphi) + a^{3}(p_{\varphi} + p_m) \,,
\label{eq10}
\end{equation}
Eq.\ (\ref{eq6}) being equivalent to $E_{\cal L} = 0$, where  $E_{\cal
L}$ is the energy. The configuration space is then given by ${\cal Q}
\equiv \{a,\varphi\}$ (the minisuperspace) and the tangent space by
${\cal T}{\cal Q} \equiv \{a,\dot{a},\varphi,\dot{\varphi}\}$, being
the {\it coordinates} associated with the pointlike Lagrangian $\cal
L$ just the scale factor $a$ and the scalar field $\varphi$, with
{\it velocities} $\dot{a}$, $\dot{\varphi}$. It is important to
stress that from Eq.\ (\ref{eq10}) we get that $E_{\cal L} =
\text{const.}$; making the homogeneous and isotropic limit of Einstein
field equations \ (\ref{eq2}) implies to choose such a constant equal
to zero. According to Noether's theorem, the existence of a symmetry
for the dynamics derived from $\cal L$ involves a constant of motion.
As a consequence, Noether symmetries in cosmology give the possibility
to infer some transformations of variables which often lead to deduce
exact cosmological solutions \cite{deR1,deR2,RivNC}. Such solutions,
though obtained by means of a procedure suggested by the existence of
this kind of symmetries, are actually independent of it and could also
be got by chance, suitably choosing the {\it right} transformation of
variables. That is, a Noether symmetry simply {\it suggests} that such
a transformation should exist and gives an easy way to find it: once we
have the {\it right} way to write down equations, we can easily solve
them and get a solution. We have to stress, also, that, in order to
verify the existence of a Noether symmetry, we find a way of
assigning the functions $F(\varphi)$ and $V(\varphi)$ for which a
Noether symmetry exists, leading to remarkable results in many cases
(see again \cite{RivNC} and references therein for several examples).
In the following, we will introduce a class of exact solutions which
deserves a special attention in the context we are working in. 

\section{A special class of solutions} 

First of all, let us notice that we start from the pointlike
Lagrangian \ (\ref{eq10}). It represents a whole class of theories,
since a particular model is assigned by specifying $F(\varphi)$ and
$V(\varphi)$. Examining the existence of Noether symmetries when the
matter content is dust ($p_m = 0$), as a first result \cite{CR2,RivNC}
one deduces the relevant relation 

\begin{equation}
V(\varphi) = V_{0} {F(\varphi)}^{p(s)} \,,
\label{eq11a}
\end{equation} 
where $V_0 > 0$ is an arbitrary constant assumed always positive, and

\begin{equation}
p(s) = \frac{3(s + 1)}{2s + 3} \,.
\label{eq11b}
\end{equation}
Therefore, the potential, through its exponent $p(s)$, depends on the
free parameter $s$. (The case $s = -3/2$ is degenerate and has been
studied separately \cite{CR3,RivNC}. When $s = -1$, $p(s)$ is zero
and $V(\varphi)$ becomes a constant; this situation has also been
treated apart \cite{RivNC}.)

The existence of such symmetries for Lagrangian \ (\ref{eq10}) implies
that we can find a differential equation for $F(\varphi)$ which has a
general solution expressed by an elliptical integral of second kind,
but we will limit our attention here to the particular solution

\begin{equation}
F(\varphi) = k_{0}{\varphi}^2 \,,
\label{eq12}
\end{equation}
where $k_{0} < 0$ is an arbitrary constant. (Negative values of
$F(\varphi)$ are necessary to disregard repulsive gravity.) Then,
Eq.\ (\ref{eq12}) inserted into Eq.\ (\ref{eq11a}) gives

\begin{equation}
V(\varphi) = \left[ V_{0} {k_0}^{p(s)} \right] {\varphi}^{2p(s)} \,.
\label{eq13} 
\end{equation}

It is possible to see that there is a whole family of exact
solutions for the time evolutions of $a$ and $\varphi$, and that they
can be expressed as \cite{RivNC}

\begin{equation}
a(\tau) = {\xi}(s) {\tau}^r \,, 
\label{eq14} 
\end{equation}

\begin{equation} 
\varphi(\tau) = {\zeta}(s) {\tau}^{6/{\chi}(s)} \,, 
\label{eq15} 
\end{equation}
where 

\begin{equation}
{\chi}(s) \equiv -\frac{6s}{2s + 3} \,,\; {\zeta}(s) \equiv \left[
\frac{{\chi}(s)}{3} \right]^{3/{\chi}(s)} \,,
\label{eq16a}
\end{equation}

\begin{equation}
{\xi}(s) \equiv {{\zeta}(s)}^{-2/[3p(s)]} \,,\; r \equiv \frac{2s^{2}
+ 9s + 6}{s(s +3)}  
\label{eq16b}
\end{equation}
are parameters depending on $s$. Time $\tau$ is, actually, a
rescaled time.

Apart from the cases showing pathologies in the solutions ($s = 0$
and $s = -3$), which have to be discussed separately \cite{CR2}, it is
important to notice that the right sign of the coupling, i.e.
$F(\varphi) < 0$, implies

\begin{equation}
-2 < s < -1 \,.
\label{eq17}
\end{equation}
When $s$ varies in such an interval, we have an infinite number of
exact solutions of the forms given in Eqs.\ (\ref{eq14}) and \
(\ref{eq15}). Asymptotically, we have

\begin{equation}
a(\tau) \approx {\tau}^{r} \,,\, {\varphi}(\tau) \approx
{\tau}^{-(2s + 3)/s} \,.  
\label{eq18} 
\end{equation} 
That is, depending on the values of $s$, the scale factor $a(\tau)$
can have asymptotic Friedmann, power--law and pole-like behaviors. 
For instance, when $|s| \gg 0$, it is $a(\tau) \approx {\tau}^{2}$. 
But, in the range of values in Eq.\ (\ref{eq17}), it can only be
Friedmannian or power--law.

As to the scalar field $\varphi$, it diverges for $s < -3/2$ and
converges for $s > -3/2$. In what follows, the range of values
anyway chosen for $s$ will always be in the latter interval.

\section{A family of exact tracker solutions}

Let us pose

\begin{equation}
\alpha \equiv - 2p(s) > 0 \,.
\label{eq19}
\end{equation}
From the definition of $p(s)$ in Eq.\ (\ref{eq11b}), this implies

\begin{equation} 
- \frac{3}{2} < s < - 1 \,,
\label{eq20} 
\end{equation}
and Eq.\ (\ref{eq13}) gives

\begin{equation} 
V(\varphi) = \left( V_{0} {k_{0}}^{-{\alpha}/2}
\right){\varphi}^{-\alpha} \,, 
\label{eq21} 
\end{equation}
with $\alpha$ always both positive and even. Substituting $\alpha$
into Eqs.\ (\ref{eq14}), \ (\ref{eq15}), \ (\ref{eq16a}), and \
(\ref{eq16b}) yields 

\begin{equation} 
a(\tau) = \left ( \frac{2 + \alpha}{3} \right )^{4/[\alpha(2 + \alpha)]}
{\tau}^{\frac{2(2{\alpha}^{2} + 9\alpha + 6)}{3({\alpha}^{2} + 6\alpha
+ 8)}} \,, \label{eq22}
\end{equation} 

\begin{equation} 
{\varphi}(\tau) = \left( \frac{2 + \alpha}{3} \right)^{3/(2 + \alpha)}
{\tau}^{2/(2 + \alpha)} \,.  
\label{eq23} 
\end{equation}

On the other hand, from Eq.\ (\ref{eq9}) we see that, being

\begin{equation} 
x \equiv \frac{{\dot{\varphi}}^{2}}{2V} > 0 
\label{eq24} 
\end{equation}
(where now dot indicates derivative with respect to rescaled time
$\tau$), we can write
 
\begin{equation}
w_{\varphi} = \frac{x - 1}{x + 1} \,.
\label{eq25}  
\end{equation}
Thus, it is clear that it is always $w_{\varphi} > -1$. For a
constant $\varphi$, i.e. a constant potential (which mimics a
cosmological constant term), we should get $w_{\varphi} = -1$. We
have in general

\begin{equation}
x \equiv \frac{{\dot{\varphi}}^{2}}{2V} =
\frac{{k_0}^{\alpha/2}}{2V_0} {\dot{\varphi}}^{2} {\varphi}^{\alpha}
> 0 \,,
\label{eq26}  
\end{equation}
implying the constraint

\begin{equation}
{k_0}^{\alpha/2} > 0 \,,
\label{eq27}  
\end{equation}
which is always true.

One of the main requests that the scalar field has to satisfy, in
order for it to be seen as a good tracker field, is that
${\dot{\varphi}}^{2} < V(\varphi)$, i.e. that $x < 1$. Requiring that
$0 < x < 1$ then poses the condition

\begin{equation}
0 < {k_0}^{\alpha/2} < \frac{27V_0}{2(2 + \alpha)} \,.
\label{eq28} 
\end{equation}
Of course, $0 < x < 1$ also implies that $w_{\varphi} < 0$, so that
it is $-1 < w_{\varphi} < 0$, namely what is needed for the scalar
field to be interpreted as quintessence. Recently, it has been
claimed that constraints from large--scale structure together with
SNIa data imply $w_{\varphi} < 0.6$ with $95\%$ of confidence level
\cite{Perl3}, which forces the field of variation for $x$ to be

\begin{equation}
0 < x < 0.25 \,.
\label{eq29} 
\end{equation}
This, in turn, yields

\begin{equation}
0 < {k_0}^{\alpha/2} < \frac{27V_0}{8(2 + \alpha)} \,.
\label{eq30} 
\end{equation}
Constraints given by Eq.\ (\ref{eq28}) are also consistent with a
direct calculation of $w_{\varphi}$ from Eq.\ (\ref{eq9}). As a
matter of fact, inserting into it the expression of Eq.\
(\ref{eq23}) for ${\varphi}(\tau)$, we get

\begin{equation}
w_{\varphi} = \frac{2{k_{0}}^{\alpha/2} (2 + \alpha) -
27V_0}{2{k_{0}}^{\alpha/2} (2 + \alpha) + 27V_0} \,.
\label{eq31}
\end{equation}
That is, $w_{\varphi} < 0$ implies Eq.\ (\ref{eq28}).

A further restriction can be found on the values of $s$, already such
that Eq.\ (\ref{eq20}) holds, if one looks at equipartition at the
end of inflation for inverse power--law potentials. This, being
$\alpha$ even, constrains to $\alpha > 5$ \cite{SWZ}, so that

\begin{equation}
- \frac{3}{2} < s < - \frac{21}{16} \,.
\label{eq32} 
\end{equation}

In \cite{SWZ}, in a minimal coupling regime, there were also
introduced two important equations, the {\it equation of motion} and
the {\it tracker equation}. As a matter of fact, the first one can
be easily generalized to the nonminimal coupling situation, giving

\begin{equation}
\pm \frac{V'}{V} = 3\sqrt{\frac{\kappa(1 +
w_{\varphi})}{{\Omega}_{\varphi}}} \left[ 1 + \frac{1}{6}\frac{d\ln
{x}}{d\ln {a}} + \frac{2F'}{\dot{\varphi}H} \left( 2H^{2} + \dot{H}
\right) \right] \,,
\label{eq33} 
\end{equation}
where $\kappa \equiv 8\pi G/3$ (we are changing now our units
following the current litereture), ${\Omega}_{\varphi} \equiv
\kappa{\rho}_{\varphi}/H^{2}$, and $F' = 2k_{0}\varphi$. (The $\pm$
signs, respectively, depend on whether $V' > 0$ or $V' < 0$.) $F' =
0$ gives the minimal coupling case, and all cosmological solutions
converge to the tracking solution, which is such that $w_{\varphi}$
is nearly constant and less than $w_{B}$ (where $B$ indicates {\it
background}), implying that $1 + w_{\varphi} = O(1)$ and therefore
${\dot{\varphi}}^{2} \approx {\Omega}_{\varphi}H^{2} = \kappa
{\rho}_{\varphi}$, so that 

\begin{equation}
\frac{V'}{V} \approx \frac{1}{\sqrt{{\Omega}_{\varphi}}} \approx
\frac{H}{\dot{\varphi}} \,. 
\label{eq34}
\end{equation}
This is referred to as the {\it tracker equation} \cite{SWZ}.

We can also introduce the function $\Gamma \equiv {V''}V/(V')^{2}$,
used in \cite{SWZ} for a test on the tracker behavior. Let us
notice, then, that we have at all times

\begin{equation} 
\Gamma \equiv \frac{V''V}{(V')^{2}} = \frac{2p(s) - 1}{2p(s)} = 1 +
\frac{1}{\alpha} = \text{const.} \,, 
\label{eq35}
\end{equation}
which is the major condition for a tracker behavior. It is $\Gamma >
1$ for $s$ in the interval in Eq.\ (\ref{eq20}), being

\begin{equation} 
\Gamma = \frac{4s + 3}{6(s + 1)} = 1 - \frac{2s + 3}{6(s + 1)} \,.
\label{eq36} 
\end{equation}
Of course, the range of values in Eq.\ (\ref{eq32}) for $\alpha > 5$
is contained in the one in Eq.\ (\ref{eq20}), still implying
therefore $\Gamma > 1$.

The tracker equation is then

\begin{eqnarray}
\Gamma = 1 + \frac{1}{(1 + w_{\varphi}) \left [
6 + \dot{x} + \frac{12F'}{\dot{\varphi}H} \left ( 2H^2 + \dot{H} \right ) \right ] } \ \nonumber \\
\times \left \{ \left [ \frac{2 \dot{x}}{\dot{\varphi}( 1+ x)^2} -
\frac{\Omega_{\varphi}'}{\Omega_{\varphi}} \right ]
\sqrt{\frac{\Omega_{\varphi}}{\kappa (1 + w_{\varphi})}} -
\frac{2}{\left [ 6 + \dot{x} + \frac{12F'}{\dot{\varphi} H} 
\left ( 2H^2 + \dot{H} \right ) \right ] } \right \} \  \nonumber \\
\times \left \{ \ddot{x} + \frac{\dot{\varphi}}{H} \frac{d}{d \varphi}
\left [ \frac{12F'}{\dot{\varphi}H} \left ( 2H^2 + \dot{H} \right )
\right ] \right \} \,,
\label{eq37}
\end{eqnarray}
giving back what can be written for the minimal coupling case (when
$F' = 0$). Here, $\dot{x} \equiv d\ln{}x/d\ln{a}$ and $\ddot{x}
\equiv d^{2}\ln{}x/d\ln{a}^{2}$.

When an inverse power--law potential is considered, as shown in
\cite{Liddle} for the minimal coupling case and in \cite{Uzan} for a
nonminimal coupling case (with $F(\varphi) \equiv \xi
{\varphi}^{2}/2$, being $\xi$ a constant, but with a slightly
different Lagrangian for $\varphi$), if the perfect fluid with
${\rho}_B \propto a^{-3(1 + w_B)}$ dominates, so that $a \propto
t^{2/[3/(1 + w_B)]}$, then the following relation holds

\begin{equation}
w_{\varphi} \approx \frac{w_{B} \alpha - 2}{\alpha + 2} \,.
\label{eq38}
\end{equation}
On the other hand, if $\Gamma$ is nearly constant, Eq.\ (\ref{eq38})
in the minimal coupling regime implies \cite{SWZ} that a solution
exists in which $w_{\varphi}$ is also nearly constant and $x$,
$\dot{x}$, $\ddot{x}$ become nearly zero. This also leads to Eq.\
(\ref{eq38}).

>From Eq.\ (\ref{eq35}), relation in Eq.\ (\ref{eq38}) can be written
as

\begin{equation}
w_{B} - w_{\varphi} = \frac{2(\Gamma - 1)(w_{B} - 1)}{1 + 2(\Gamma
- 1)} \,,
\label{eq39} 
\end{equation}
evidentiating the fact that $\Gamma > 1$ is equivalent to $w_{B} >
w_{\varphi}$ in a matter dominated situation.

In \cite{Uzan} it is found a solution for $\varphi$ of the same type
as in Eq.\ (\ref{eq23}). Thus, the situation described therein is
practically similar to ours, and we can import some of its
considerations. Assuming that the universe is matter dominated, $a
\propto t^{2/[3/(1 + w_B)]}$ implies that $H = [2/3(1 + w_{B})]
t^{-1}$, and the Klein--Gordon equation \ (\ref{eq5}) can be written
as

\begin{eqnarray}
\ddot{\varphi} + \frac{2}{1 + w_{B}} \frac{1}{\tau}\dot{\varphi} -
\frac{8}{(1 + w_{B})} \left[ \frac{4}{3(1 + w_{B})} - 1 \right]
k_{0}\frac{1}{{\tau}^{2}}\varphi \nonumber \\
- \alpha V_{0} k^{-\alpha/2}_{0} {\varphi}^{-\alpha -1} = 0 \,.
\label{eq40}
\end{eqnarray}
As demonstrated in \cite{Uzan}, this involves, for example, that we
can take Eq.\ (\ref{eq38}) as valid also in our context, even if we
did not write Eq.\ (\ref{eq37}) in such a way to clearly evidentiate
that behavior when $x$,$\dot{x}$, $\ddot{x}$ are negligible and
$w_{\varphi}$ is constant.

If we, only in a speculative way, compare Eq.\ (\ref{eq31}),
``exact'' and always valid at any time, and Eq.\ (\ref{eq38}),
``approximated'' and valid only when ${\rho}_{B} \gg
{\rho}_{\varphi}$, we find

\begin{equation}
{k_{0}}^{\alpha/2} = \frac{\alpha (1 + w_{B})}{4(2 + \alpha)} \,,\,
V_{0} = \frac{4 + \alpha (1 - w_{B})}{54} \,.
\label{eq41}
\end{equation}
This forces the constant ${k_{0}}^{\alpha/2}$ to be

\begin{equation}
{k_{0}}^{\alpha/2} = \frac{1}{2} - \frac{27 V_{0}}{2(2 + \alpha)}
\label{eq42}
\end{equation}
and, according to Eq.\ (\ref{eq28}), leads to 

\begin{equation}
\frac{2 + \alpha}{54} < V_{0} < \frac{2 + \alpha}{27}
\label{eq43}
\end{equation}
or, from Eq.\ (\ref{eq30}), to

\begin{equation}
\frac{4(2 + \alpha)}{135} < V_{0} < \frac{2 + \alpha}{27} \,. 
\label{eq44} 
\end{equation}
Also, substituting Eq.\ (\ref{eq41}) directly into Eq.\ (\ref{eq28})
and Eq.\ (\ref{eq30}), respectively, gives

\begin{equation}
w_{B} < \frac{2}{\alpha} \;,\; w_{B} < \frac{4 - 3\alpha}{5\alpha}
\label{eq45}
\end{equation}
or, equivalently,

\begin{equation}
\alpha < \frac{2}{w_{B}} \;,\; \alpha < \frac{4}{5w_{B} + 3} \,.
\label{eq46}
\end{equation}
Let us notice, then, that the second relation in Eq.\ (\ref{eq45})
always implies $w_{B} < 0$ (which is not good for ordinary matter),
while the first relation in Eq.\ (\ref{eq45}) gives the {\it right}
constraint $w_{B} < 1$. Also, the second relation in Eq.\
(\ref{eq46}) yields $\alpha < 4/3$ for dust ($w_{B} = 0$) and $\alpha
< 1/2$ for stiff matter ($w_{B} = 1$), while the first relation in Eq.\
(\ref{eq46}) lets us accept any value of $\alpha$ for dust, and gives
$\alpha < 6$ for radiation ($w_{B} = 1/3$) or $\alpha < 2$ for stiff
matter. Finally, we can say that these considerations seem to imply
that Eq.\ (\ref{eq29}) limits too much the variability of $x$.

Let us also notice that Eq.\ (\ref{eq38}) could be read as

\begin{equation}
w_{B} \approx \frac{2 - (2 + \alpha)|w_{\varphi}|}{\alpha} \,.
\label{eq47}
\end{equation}
Now, for ordinary matter (i.e., when $0 \leq w_{B} \leq 1$), it
comes out (always disregarding the approximated equality)

\begin{equation}
\frac{2 - \alpha}{2 + \alpha} \leq |w_{\varphi}| \leq \frac{2}{2 +
\alpha} \,,
\label{eq48} 
\end{equation}
implying (for $\alpha = 2$) $0 \leq w_{\varphi} \leq 1/2$, or (for
$\alpha = 4$) $0 < |w_{\varphi}| \leq 1/3$, for instance. This means,
then, that values $w_{\varphi} \lesssim 0.6$ are however possible,
even if considerations made above let us understand that there may
also be values of $w_{\varphi}$ such that $0.6 < w_{\varphi} < 0$.

Of course, all these kinds of considerations have to be taken just as
indicative, since Eq.\ (\ref{eq38}) is not always valid and is
usually read in one way: once we assign a specific value of
$w_{B}$, then it gives an approximated value of $w_{\varphi}$. For
example, $w_{B} = 0$ gives $w_{\varphi} \approx -2/(2 + \alpha) < 0$
for any $\alpha > 0$, and from $w_{B} = 1/3$ it is found $w_{\varphi}
\approx (\alpha - 6)/[3(2 + \alpha)]$ (giving the critical value
$\alpha = 6$, such that $\alpha > 6$ implies positive $w_{\varphi}$
and $\alpha > 6$ implies negative $w_{\varphi}$).

On the other hand, anyway, taking the tracker condition in Eq.\
(\ref{eq34}) into account, we can immediately control its validity,
being

\begin{equation} 
\frac{V'}{V} = - \frac{\alpha}{\varphi} \approx {\tau}^{\frac{1 +
w_{\varphi}}{1 + w_{B}} - 1} \,,
\label{eq49a} 
\end{equation}

\begin{equation}
\frac{H}{\dot{\varphi}} \approx {\tau}^{\frac{1 + w_{\varphi}}{1 +
w_{B}} - 1} \,.
\label{eq49b}
\end{equation}
As a global consequence, we can assert that \ (\ref{eq23}) is a
{\it good} family of exact tracker solutions, parametrized by the
constants $V_{0}$ and $k_{0}$, being

\begin{equation}
M^{4 + \alpha} \equiv V_{0}k^{-\alpha/2}_{0} \,.
\label{eq50}
\end{equation}
If we consider relations in Eq.\ (\ref{eq41}), apart from the
contribution of the ordinary matter content, the parameter $M$
depends only on $s$

\begin{equation}
M = \left\{ \frac{4s \left[ s \left( \frac{1 + 3w_{B}}{1 + w_{B}}
\right) + 3 \right]}{81(s + 1)(2s + 3)} \right\} ^{(2s + 3)/[2(s +
3)]} \,.
\label{eq51}
\end{equation}
The observational constraint ${\Omega}_{\varphi} \approx 0.7$ today,
when $\varphi \approx O(M_{P})$, implies that $V(\varphi \approx
M_{P}) \approx {\rho}_{m_{0}}$, being ${\rho}_{m_{0}} \approx 10^{-47}
\text{GeV}^{4}$ the current matter density. This gives \cite{ZWS}

\begin{equation}
V(\varphi) = M^{4 + \alpha}{\varphi}^{- \alpha} \approx M^{4 +
\alpha}M^{- \alpha}_{P} \approx {\rho}_{m_{0}} \,,
\label{eq52}
\end{equation}
so that $M \approx ({\rho}_{m_{0}}M^{- \alpha}_{P})^{1/(4 + \alpha)}
> 1 \text{GeV}$, in good comparison to particle physics scale, when
$\alpha \gtrsim 2$, that is $s \lesssim -1.2$. (This shows again
that, since such a value does not respect Eq.\ (\ref{eq32}),
constraining to $\alpha > 5$ should be reconsidered.)

\section{Conclusions}

In the picture offered by observations in recent times, a good
cosmological scenario needs an energy component with negative
pressure. Of course, the simplest and most extreme candidate is a
cosmological constant, but other {\it softer} proposals exist. Among
these, quintessence has been advanced resuming, generalizing, and
suitably readapting older ideas on cosmology with a scalar field.
Initially thought of as a field acting like an attractor on other
solutions, a more refined version of it has been proposed more
recently, namely the tracker field. This kind of field tends to
isotropize the universe at late times, nicely solving the coincidence
problem. In that, it finds an ideal convergence with what is claimed by
Wald's theorem on isotropization in homogeneous cosmologies with a
positive cosmological constant. 

In this paper, we have shown that the requested features of such a
field when $V(\varphi)$ has an inverse power--law behavior can be
obtained looking at a kind of exact solutions already present in the
past literature on nonminimally coupled scalar--tensor theories of
gravity (see \cite{RivNC}, for example). Even if it was deduced in a
very peculiar way there, nevertheless it has been reintroduced as,
say, an {\it ad hoc} tracker solution only more recently. For these
reasons, it has seemed interesting to connect the solution for
$\varphi$ to the solution for $a$, placing all the discussion in the
context of a well developed theory. To be precise, then, a family of
exact tracker solutions has been found here, depending on the values
of a parameter $s$ which is crucial for the relationship existing
between the coupling $F(\varphi)$ and the potential $V(\varphi)$ in
our context.

Such a relationship is a condition for a Noether symmetry to exist
in the cosmological scenario involved by the model proposed through
action in Eq.\ (\ref{eq1}). This condition appears to be very
important to us, because it seems to imply tracker fields in a very
natural way, based on the apparent naturality of Noether symmetries
in cosmology \cite{RivNC}.

Also, in the context of nonminimally coupled scalar--tensor theories
of gravity, there is no dramatic difference between quintessence and
cosmological constant proposals, introduced to solve the puzzle
offered by recent observations. It can be shown that there is, in
fact, an evident possibility to generalize Wald's theorem \cite{Wald}
in order to get an asymptotic cosmological constant in many
nonminimal theories, without introducing it {\it a priori} (see
\cite{CR4} and references therein). As a matter of fact, for many
choices of $F(\varphi)$, a time dependent $\Lambda$--term can be
defined, asymptotically approaching a constant (that is, a {\it
cosmological} constant). (In \cite{CR4} several examples are
considered.) In this connection, it is noteworthy that some
interesting comments have also been made \cite{MPR} on the fact that,
with respect to the coincidence problem, an inverse power--law
potential is not really different from a cosmological constant. 
 
\section*{Acknowledgments}

This work has been finantially sustained by the M.U.R.S.T. grant
PRIN97 ``SIN.TE.SI.''.

\end{document}